\documentclass[journal=apchd5,manuscript=article]{achemso}

\usepackage[version=3]{mhchem} 
\usepackage{xcolor}
\usepackage{textcomp}
\usepackage{gensymb}
\usepackage{verbatim}

\author{Prince Khatri}
\affiliation{College of Engineering, Mathematics and Physical Sciences,University of Exeter, Exeter EX4 4QF, United Kingdom}
\author{Ralph Nicholas Edward Malein}
\affiliation{College of Engineering, Mathematics and Physical Sciences,University of Exeter, Exeter EX4 4QF, United Kingdom}
\author{Andrew J. Ramsay}
\affiliation{Hitachi Cambridge Laboratory, Hitachi Europe Ltd., Cambridge CB3 0HE, United Kingdom}
\author{Isaac J. Luxmoore}
\affiliation{College of Engineering, Mathematics and Physical Sciences,University of Exeter, Exeter EX4 4QF, United Kingdom}
\email{i.j.luxmoore@exeter.ac.uk}

\title
  {Stimulated Emission Depletion Microscopy with Color Centers in Hexagonal Boron Nitride}

\begin{document}

\begin{abstract}
Stimulated emission depletion, or STED microscopy is a well-established super-resolution technique, but is ultimately limited by the chosen flourophore. Here we demonstrate STED microscopy with color centers in nanoscale flakes of hexagonal boron nitride using time-gated continuous wave STED. For color centers with zero phonon line emission around 580 nm we measure a STED cross-section of (5.5 $\pm$ 3.2) x $10^{-17} \mbox{cm}^{2}$, achieve a resolution of $\sim$ 50 nm and resolve two color centers separated by 250 nm, which is less than the diffraction limit. The achieved resolution is limited by the numerical aperture of the objective lens (0.8) and the available laser power, and we predict that a resolution of sub-10 nm can be achieved with an oil immersion objective lens, similar to state-of-the-art resolution obtained with nitrogen vacancy centers in diamond. 

\end{abstract}


Stimulated emission depletion (STED) microscopy\cite{Hell1994} is one of a number of super-resolution imaging techniques and employs a combination of two lasers: one that excites fluorescence from a flourophore, and a second that deactivates fluorescence via stimulated emission, to effectively reduce the excitation area below the diffraction limit\cite{Klar1999SubdiffractionMicroscopy}. The resolution can theoretically reach the atomic scale, but is ultimately limited by the properties of the flourophore \cite{Westphal2005NanoscaleMicroscope}. It is therefore important to investigate and characterize potential new flourophores. The lateral resolution, $\delta d \propto \sqrt{I_{sat}/I}$, where $I$ is the optical intensity used to stimulate emission and $I_{sat}=h\nu/\sigma_s\tau_{r}$ is the stimulated-emission saturation intensity of the emitter, with $h\nu$ the photon energy, $\tau_r$ the excited state radiative lifetime and $\sigma_s$ is the STED cross section for stimulated emission \cite{Vicidomini2011}. The ideal flourophore should therefore have large $\sigma_s$ to enable high resolution, whilst minimising the chance of photodamage to the sample. Flourophores should also be bright for a low excitation power, photostable, in terms of blinking and bleaching, and physically small.

The most widely used flourophores are organic dye molecules, which can have a relatively large STED cross-section on the order of $10^{-16} \mbox{ cm}^{2}$\cite{Kastrup2004}. They can be designed to bind to certain biological \cite{Sameiro2009FluorescentProbes} targets and produce high emission rates \cite{Dempsey2011EvaluationImaging} at sufficiently low power \cite{Bouzin2013StimulatedSpectroscopy} to enable spatial resolution of $\sim20$ nm \cite{Donnert2006Macromolecular-scaleMicroscopy}. However, photobleaching is a significant limitation of these organic dyes, which has stimulated the search for more stable flourophores \cite{Eggeling1998PhotobleachingPhotolysis,Oracz2017PhotobleachingFluorophore}. A prime example is the negatively charged nitrogen vacancy (NV) center in diamond, where a resolution of 2.4 nm has been demonstrated in bulk diamond\cite{Wildanger2012} and individual NV centers separated by just 15 nm have been resolved in nanodiamonds\cite{Arroyo-Camejo2013}. Nanodiamonds can be functionalized to bind with biological targets and their relatively non-toxic nature allow their use for target labelling and imaging in living tissues \cite{Man2013}. However, the STED cross-section is smaller than that of dye molecules, at around $10^{-17} \mbox{cm}^{2}$, requiring a larger depletion power \cite{Han2009Three-dimensionalLight,Rittweger2009STEDResolution}. Furthermore, the photoluminescence intensity is low and spans a large spectral range due to phonon assisted emission, and photocharging can result in blinking, particularly in nano-diamonds with diameters $<10$ nm \cite{Rabeau2007SingleNanocrystals}. In recent work, negatively charged silicon vacancies (SiV) have been proposed as an alternative\cite{Silani2019}, offering the favourable properties of diamond, but with a larger STED cross-section ($4\times10^{-17} \mbox{ cm}^{-2}$). 

These impressive results with diamond raise the question of whether color centers in other wide bandgap semiconductors have potential for STED microscopy. In particular, color centers in hexagonal boron nitride appear an interesting candidate, thanks to their optical properties and the 2D nature of the host material \cite{Tran2016QuantumMonolayers}. Stable emission has been demonstrated from hBN nanoflakes with thicknesses of just a few atomic layers and lateral dimensions $<10$ nm \cite{Duong2019}. Colour centers have been reported with emission energy spanning the UV to infra-red\cite{Bourrellier2016BrightH-BN,Tran2016RobustNitride,Camphausen2020ObservationTemperature,Vuong2016a,Konthasinghe2019RabiEmitter}, with the emission often concentrated in a narrow and bright zero phonon line (ZPL). Color centers in hBN have previously been studied with super-resolution microscopy\cite{Kianinia2018,Feng2018a}, and stimulated depletion has been employed as spectroscopic probe of electron-phonon coupling \cite{Malein}, but they have not previously been employed in STED microscopy.

In this work, we demonstrate STED microscopy using single color center defects in hBN flakes. For color centers emitting around 580 nm we find an average STED cross-section of $\sigma_{S} = (5.5\pm3.2) \times 10^{-17} \mbox{ cm}^2$, and up to $\sim1\times10^{-16} \mbox{ cm}^2$ in some defects. We demonstrate a spatial resolution of approximately 50 nm, limited by the modest numerical aperture (0.8) of our system, and resolve two color centers separated by less than the diffraction limit.    

\begin{figure}
\begin{center}
\vspace{0.2 cm}
\includegraphics[width=1\columnwidth]{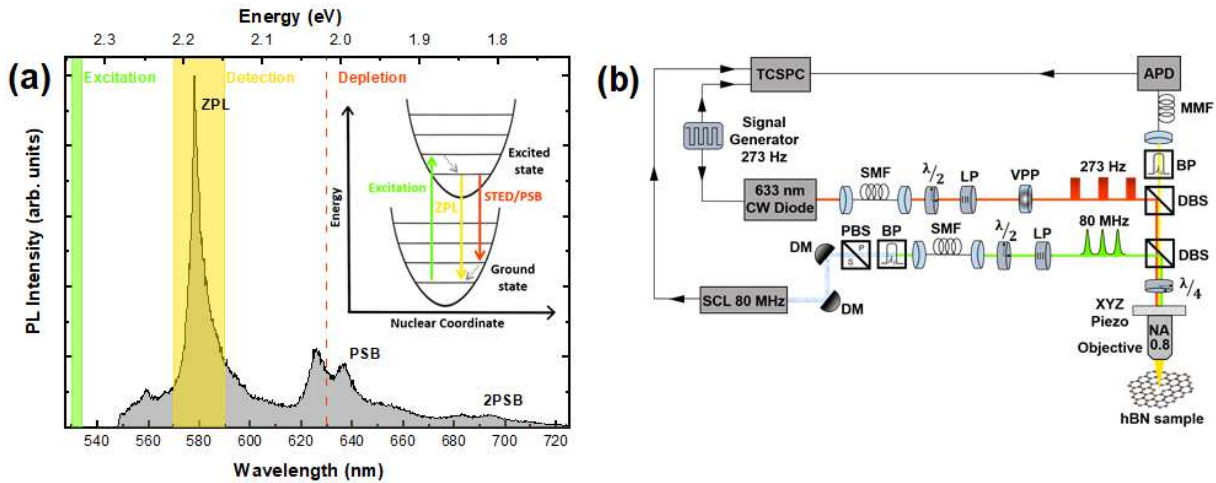}
\end{center}
\caption{   
(a) Typical PL spectrum of a color center in hBN, showing the excitation, depletion and collection wavelengths. The labels ZPL, PSB and 2PSB indicate the zero-phonon line, one optical phonon sideband and two optical phonon sidebands, respectively. (b) Schematic diagram of the gated CW-STED setup. DM=dielectric mirror; SMF=single-mode fiber; MMF=multi-mode fiber PBS=polarizing beam-splitter; BPF=band-pass filter; VPP=vortex phase plate; LP= linear polarizer; DBS=dichroic beam-splitter; SCL=supercontinuum laser; APD=avalanche photodiode; $\lambda/4$=quarter-waveplate; $\lambda/2$=half-waveplate; TCSPC-time-correlated single photon counting electronics.}
\label{fig1}
\end{figure}

The sample consists of few-layer flakes of hBN drop-cast from a water-ethanol solution onto a silicon substrate coated with 5 nm of $\mbox{Al}_{2}\mbox{O}_{3}$. In this work we focus on a class of defects with zero-phonon line (ZPL) emission at around 580 nm. Fig. \ref{fig1}(a) shows the photoluminescence (PL) spectrum from such a defect, under green (532nm) excitation. As shown in the inset to Fig. \ref{fig1}(a), following green excitation and fast relaxation to the excited state, the defect can relax directly via the zero phonon line, or via a phonon-assisted transition, with the one and two optical phonon sidebands (OPSB) clearly visible in \ref{fig1}(a), with detuning of $\sim{-175}$ meV and $\sim-350$ meV, respectively. A red laser (630 nm), with doughnut-shaped intensity profile, stimulates phonon-assisted emission, thereby reducing the ZPL luminescence.

The experimental setup for STED-imaging is shown in Fig. \ref{fig1}(a) and employs a variant of STED known as time-gated CW-STED \cite{Vicidomini2011}, where the excitation (depletion) laser is pulsed (continuous-wave [CW]). The excitation is generated by a supercontinuum laser, with a repetition rate of 78 MHz that can be reduced using a built-in pulse-picker. The supercontinuum beam is directed by a pair of dielectric mirrors, which filter-out wavelengths $>$800 nm, through a polarizing beam-splitter and band-pass filter centerd at 532 nm (4 nm FWHM), and into a single mode fiber. At the output of the fiber a half-wave plate and linear polarizer maximize the power and ensure a high degree of linear polarization, respectively. The depletion beam is provided by a red CW diode laser, with a wavelength of 630 and is modulated on and off at 273 Hz. The red laser is coupled to a single mode fiber, followed by a half-wave plate, linear polarizer and vortex phase plate, which generates the doughnut shaped beam profile necessary for STED microscopy \cite{Hao2010EffectsMicroscopy}. Two dichroic beam splitters combine the green and red lasers, which are passed to an objective lens with numerical aperture of 0.8 that is mounted on an XYZ-piezo stage. The fluorescence from the sample is collected by the same objective, filtered by a pair of tunable long and short pass filters (570-590 nm pass band) and coupled, via a multi-mode fiber, into a single photon avalanche photodiode (SPAD) module. A time correlated single photon counting (TCSPC) module records the arrival time of photons, along with reference pulses from the two lasers. This data is then post-processed to generate time-resolved decay traces and confocal/STED maps. Modulation of the depletion laser is applied so that confocal and STED maps can be acquired simultaneously. Where quoted, the laser power is measured before the objective lens and represents an upper limit, as both excitation and depletion lasers are collimated to a diameter slightly larger than the back aperture of the objective lens.     

\begin{figure}
\begin{center}
\vspace{0.2 cm}
\includegraphics[width=1\columnwidth]{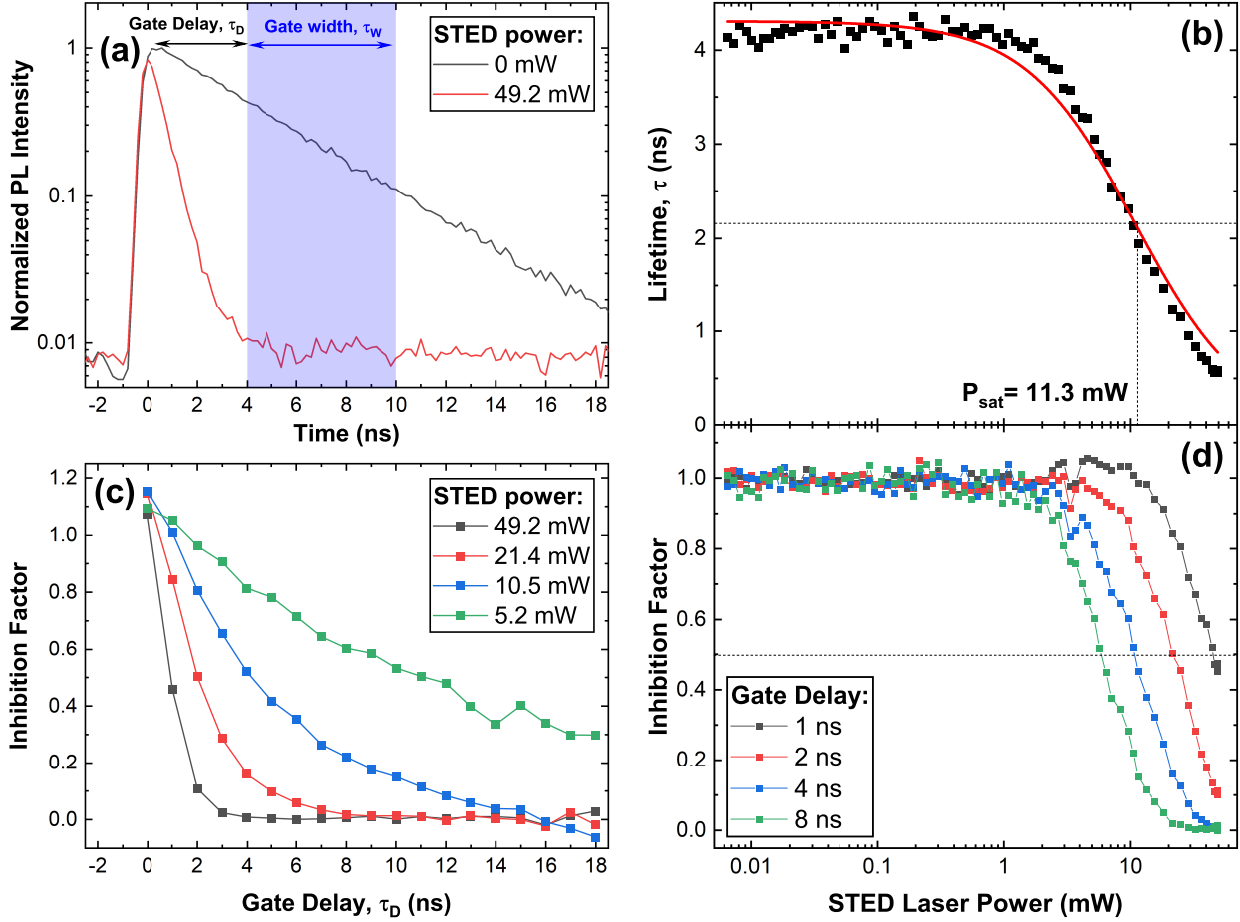}
\vspace{0.2 cm}
\end{center}
\caption{Time-gated CW-STED with a color center in hBN. (a) Time-resolved photoluminescence from a single hBN color center, under pulsed 532nm excitation, with and without 633nm CW-STED laser co-excitation (b) STED laser power dependence of the radiative lifetime of the colour center. The red line shows a fit to eq. \ref{eqn1}. (c) and (d) STED inhibition factor as a function of (c) gate delay time and (d) STED laser power with gate width, $\tau_w =$ 6 ns. The depletion laser has a Gaussian-shaped intensity profile.}
\label{fig2}
\end{figure}

In general, using a CW rather than pulsed STED laser has the advantages of lower cost and ease of implementation, but requires much higher average power to achieve the same resolution \cite{Willig2007STEDBeams}. Time-gated CW-STED, which is demonstrated in Fig. \ref{fig2}(a) for an hBN color center using a Gaussian-shaped depletion beam, overcomes this problem by combining a CW depletion laser with pulsed excitation and time-gated detection, and can yield high resolution at low STED intensities\cite{Vicidomini2011}. In this approach, the lifetime is reduced by the depletion laser from $1/k_{fl}$ to:

\begin{equation}
   \tau=\frac{1}{k_{fl}+(P_{S}\sigma_{S}\lambda_{S} \big/ A_{S}hc)} \label{eqn1}
\end{equation}

where $k_{fl}$ is the radiative decay rate of the flourophore, $P_{S}$ the power of the STED laser of wavelength $\lambda_{S}$, $A_{S}=2\pi (SD)^2$ is the STED beam focal area. $A_S$ is approximated by the standard deviation ($SD = 246 \ \textrm{nm}$) of the Gaussian profile of the focal spot without VPP in beam path (see Supplementary material [S.M.] for details). Collecting only the photons emitted  after the CW laser has depleted the excited state, $\tau_d>\tau$, effectively reduces the STED inhibition factor (the ratio of fluorescence intensity with and without the depletion laser applied), and therefore increases the STED resolution. This enables lower depletion laser power to be compensated by a longer gate delay, as illustrated in Fig. \ref{fig2}(c). The trade-off is that with longer gate delays the number of photons in the collection window is reduced, thereby reducing the signal to noise ratio. To some extent this can be compensated by choosing a gate width, $\tau_W$ and excitation laser repetition rate, that maximises the fluorescence intensity based on the radiative lifetime of the emitter.

In Fig \ref{fig2}.(b) the emitter lifetime is plotted as a function of the depletion laser power and is reduced from $\sim4$ ns to $\sim0.5$ ns for a STED power, $P_{S} = 49  \textrm{ mW}$, which is the maximum available from this laser. The STED cross-section is extracted from a fit to eq.~\ref{eqn1} 
and for this color center yields a value of  $\sigma_{S} \geq 2.5\times 10^{-17} \mbox{cm}^2$ (see S.M. for details). Note, as the power is measured before the objective lens, the power incident on the sample is less than quoted and hence our calculation using eq. \ref{eqn1} gives a lower limit of the STED cross section. Increasing the time delay of the gate reduces the STED saturation power, defined as the power required to reduced the fluorescence intensity by half,  (Fig. \ref{fig2}(d)). The measurements shown in Fig. \ref{fig2} are repeated for ten more color centers resulting in a mean STED cross-section of $ \geq (5.5\pm3.2) \times 10^{-17} \mbox{cm}^2$, with a maximum value of $\geq 1\times10^{-16} \mbox{cm}^2$ (see S. M.). These cross-sections compare favourably with centers in diamond \cite{Silani2019,Rittweger2009STEDResolution} and organic dye molecules\cite{Kastrup2004,Vicidomini2011} and demonstrate the clear potential for hBN color centers to find application in STED microscopy. The relatively large variation in the STED cross-section can be attributed to variations in the efficiency of the optical phonon-assisted depletion from defect to defect. Unlike for NV and SiV centers in diamond, where the ZPL wavelength is well defined, in hBN samples such as ours, there is considerable variation of the ZPL wavelength\cite{Tran2016RobustNitride}. In our experiment, the detection bandwidth spans 20 nm, but the STED laser bandwidth is $\sim1$ nm. Consequently, the overlap of the STED laser with peaks in the OPSB can vary considerably, leading to the observed defect-to-defect variation in STED cross-section. This could be avoided by using a broadband STED laser, such as a filtered supercontinuum\cite{Silani2019}.

\begin{figure}
\begin{center}
\vspace{0.2 cm}
\includegraphics[width=1.0\columnwidth]{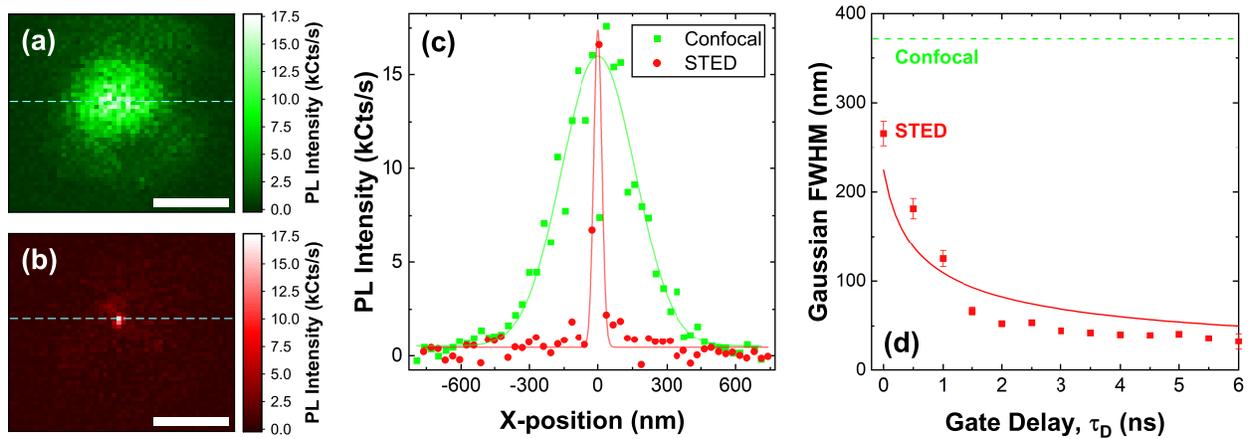}
\vspace{0.2 cm}
\end{center}
\caption{ 
STED microscopy of a single color center in hBN. (a) Confocal and (b) STED images of a single hBN color center, for a gate delay of 3 ns and a gate width of 6 ns. In (a) and (b) the scale marker is 500 nm. In (b) the STED power is 49 mW. (c) Horizontal intensity profile through the center of the confocal and STED images. The solid red and green lines show Gaussian fits to the experimental data with FWHM of 380 nm and 41 nm, respectively. (d) Gaussian FWHM of the STED image as function of gate delay time. The solid red line is a fit to the FWHM of the calculated point spread function for time-gate CW-STED (See. S. M). }
\label{fig3}
\end{figure}

To verify that hBN color centers can be applied to STED microscopy, we use a vortex phase plate to generate a doughnut-shaped depletion laser beam with a central minima co-aligned with the maxima of the Gaussian excitation beam (see S. M. for laser beam profile measurements). Scanning confocal and STED images of a single color center, presented in Fig. \ref{fig3}(a) and (b), respectively,  are acquired simultaneously by switching the depletion laser on and off at 237 Hz ($P_{S} = 49 \ \textrm{mW}$, $\tau_D=3$ ns). The images reveal a clear improvement in resolution, which is quantified with Gaussian fits to horizontal intensity profiles through the two images (Fig. \ref{fig3}(c)). The confocal FWHM is $\sim$380 nm, consistent with the diffraction limit of our microscope ($\lambda=532$ nm, $N.A.=0.8$). Whereas, with the STED laser applied the FWHM is $\sim$40 nm, which is  a $\sim$ 9-fold improvement in resolution. 

Fig. \ref{fig3}(d) highlights the value of time-gating in CW-STED, with the Gaussian FWHM plotted as a function of the gate delay time. The confocal resolution is independent of the gate delay, whereas the STED FWHM reduces from 270 nm, with no gate delay, to $\sim$40 nm when the gate delay is $>\sim2 ns$. The solid line shows a fit to the data of the calculated point spread function for time-gated CW-STED\cite{Vicidomini2011} (see S. M. for further details). On closer inspection, the STED image reveals some asymmetry, with a FWHM of $\simeq 60$ nm in the y-direction, which we attribute to aberrations\cite{Li2018EffectsMicroscopy,Antonello2017AberrationsMicroscopy} and/or drift in our setup (see S. M.). Note, the confocal FWHM is also larger in the y-direction at $\sim450$ nm. 

\begin{figure}
\begin{center}
\vspace{0.2 cm}
\includegraphics[width=0.6\columnwidth]{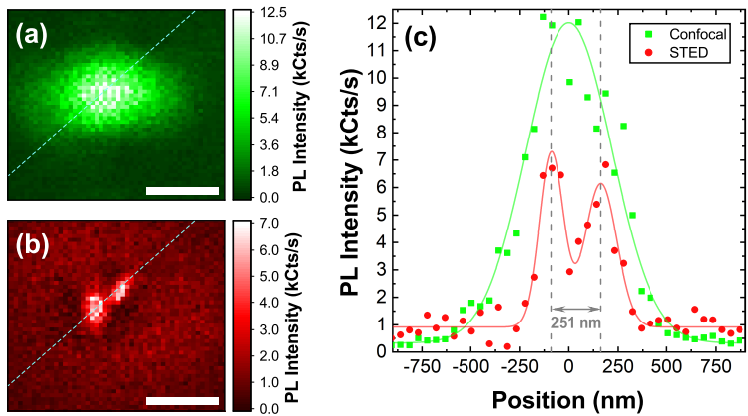}
\vspace{0.2 cm}
\end{center}
\caption{ 
(a) Confocal and (b) STED images of a multiple hBN colour centers, for a gate delay of 3 ns. In (a) and (b) the scale marker is 500 nm and the blue dashed lines indicate the linecuts shown in (c). In (b) the STED power is 49 mW. (c) and (d) Linecut intensity profiles from the (c) confocal and (d) STED images along the direction indicated in (a) and (b). The solid green (red) lines in (c) and (d) show Gaussian (double Gaussian) fits to the experimental data.}
\label{fig4}
\end{figure}

Finally, in Fig. \ref{fig4} we demonstrate that two hBN color centers, separated by less than the diffraction limit, can be resolved using STED microscopy. In Fig. \ref{fig4}(a) and (b), confocal and STED images of the same area of the sample are presented. In the confocal image (Fig. \ref{fig4}(a)) a single bright spot is observed, apparently indicating emission from a single color center. However, the STED image (Fig. \ref{fig4}(b)) clearly reveals two distinct emission centers. From the intensity profiles presented in Fig. \ref{fig4}(c) and (d), and Gaussian fits to the data, the separation of the two colour centers is $251\pm8$ nm, compared to the diffraction limit of $\lambda/2NA=333$ nm for our system.

In conclusion, these results clearly illustrate the potential of hBN color centers as a flourophore for STED microscopy. In our current set-up the resolution is limited by both the numerical aperture of the objective lens and the available STED laser power, yet a resolution of $\sim50$ nm is still achieved for a power density of $13 \ \textrm{MW/cm}^{2} $. With an oil-immersion lens, widely used in STED microscopy, the numerical aperture is typically 1.4, which would translate to a resolution of $<$10 nm (see S. M.), comparable to state of the art measurements with diamond NV centers\cite{Arroyo-Camejo2013}. In this work, we have employed color centers with ZPL emission around 580 nm. However, the wide range of emission energies can provide flexibility in the choice of excitation/depletion wavelength and color centers emitting at shorter wavelengths \cite{Shevitski2019,Bourrellier2016BrightH-BN} offer the potential of further improvements in resolution. Photoluminescence from sub-10 nm diameter flakes \cite{Duong2019} and the non-toxic nature of hBN\cite{Merlo2018} show promise for biological applications, but further work is required to understand the role of optical pumping on the PL stability \cite{Khatri2020,Kianinia2018,White2020OpticalNitride} so that the excitation scheme can be optimised, and to determine the underlying defect structure so that samples with a single dominant defect species can be produced \cite{Mendelson2020IdentifyingNitride}. Furthermore, STED microscopy can be combined with ODMR \cite{Wildanger2012, Wildanger2011DiffractionResonances, Arroyo-Camejo2013} to study the spin physics of color centers in hBN \cite{Gottscholl2020InitializationTemperature,Chejanovsky2019}, where there is considerable promise for applications in quantum sensing\cite{Petrini2020}.

\begin{acknowledgement}

This work was supported by the Engineering and Physical Sciences Research Council [Grant numbers EP/S001557/1 and EP/026656/1]. We thank A. Corbett, C. Soeller and B. Patton for helpful discussions. Data underlying the results presented in this paper are not publicly available at this time but may be obtained from the authors upon reasonable request.

\end{acknowledgement}

\begin{suppinfo}

\section{1. STED Beam Characterization}
To quantify the STED cross-section, the intensity profile of the laser at the focal plane is required. To measure this in our setup, we take advantage of the fact that color centers emitting over a broad energy range can be found in hBN\cite{Bourrellier2016BrightH-BN,Tran2016RobustNitride,Camphausen2020ObservationTemperature,Vuong2016a,Konthasinghe2019RabiEmitter}. Using the same sample, we locate a color center that can be efficiently excited with the red laser, in this case with ZPL emission at 685 nm, which we then use as an effective point source for mapping the STED laser beam profile. The spectrum of the color center is shown in Fig. \ref{fgr_SI_STED_beam}(a). The ZPL fluorescence is recorded whilst the beam is scanned over the defect, revealing a Gaussian beam profile, with a FWHM of 580 nm (Fig. \ref{fgr_SI_STED_beam}(b) and (d)).

\begin{figure}
  \includegraphics[width=\textwidth]{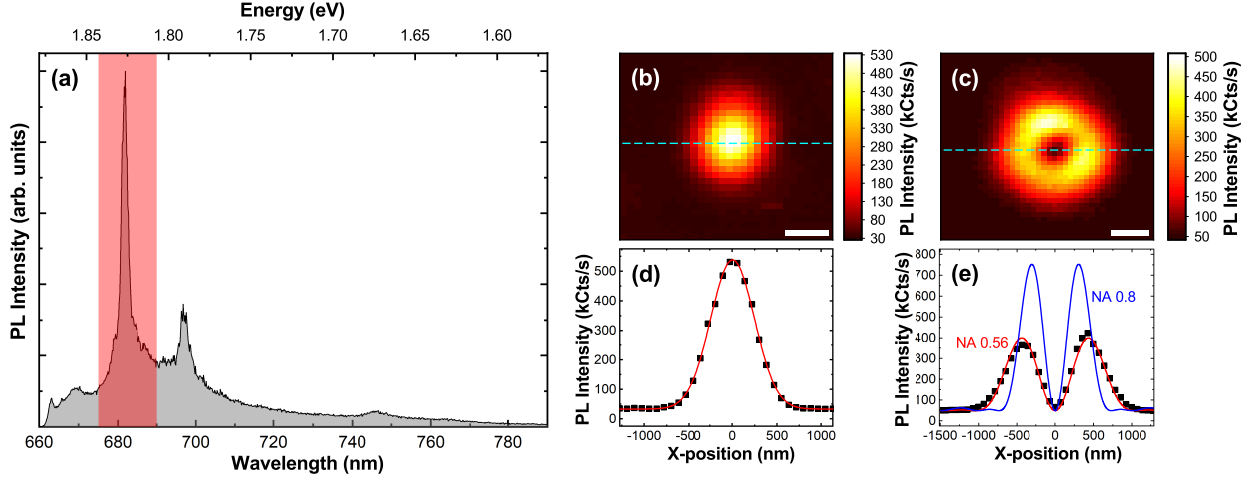}
  \caption{Characterization of STED Laser Beam. (a) PL spectrum of the hBN color center used for STED beam characterization. The shaded area indicates the wavelength range used for in the beam characterization. (b) and (c) Confocal images of the hBN color center when excited with (b) a Gaussian beam profile and (c) the doughnut beam profile. The scale marker in (b) and (c) is 500 nm and the dashed lines indicate the linecuts shown in (d) and (e). (d) and (e) PL intensity profiles from (d) the gaussian beam and (e) the doughnut beam maps. The solid red line in (d) shows a Gaussian fit to the data with FWHM of 580 nm and in (e) shows a fit to doughnut beam profile calculated with eq. \ref{Ir} giving $\textrm{NA}_{eff} = 0.56$.}
  \label{fgr_SI_STED_beam}
\end{figure}

This method is also used to measure the doughnut beam intensity profile, which is required to interpret the achieved  STED resolution, and is plotted in Fig. \ref{fgr_SI_STED_beam}(c) and (e). The doughnut beam profile, $D(r)$, along the radius, $r$, is approximated as\cite{Silani2019,Khonina1992}: 

\begin{align} \label{Ir}
D(r) \simeq \frac{2\pi \textrm{NA}^2}{\lambda_{S}^2} \left( \pi \frac{H_o(u)J_1(u) - H_1(u)J_o(u)}{u}  \right)^2 \tag{S1}
\end{align}

where NA is the numerical aperture of the objective lens, $J_o$ and $J_1$ are the zeroth and first order Bessel Functions, and $H_o$ and $H_1$ are the zeroth and first order Struve Functions with normalized radius $u=2\pi r\textrm{NA}/\lambda_{S}$.

In Fig. \ref{fgr_SI_STED_beam}(e) the experimental data is shown alongside a fit to the data of eq. \ref{Ir}, from which we extract an effective numerical aperture of our microscope, $\textrm{NA}_{eff}$ = 0.56, that is less than the the actual NA of the objective lens (0.8). This reduced NA can be attributed to imperfect circular polarization and the alignment of the doughnut beam into the objective lens. For comparison, eq. \ref{Ir} is also plotted for NA=0.8, which illustrates that larger NA leads to an increase in the intensity and reduction of the distance from the doughnut center of the two maxima.

\section{2. STED doughnut beam quality and resolution}

In Vicidomini \textit{et al.}\cite{Vicidomini2011}, an analytical expression for the point spread function (PSF) is derived for time-gated CW-STED microscopy. We adapt this approach to take into account the gate width, which in this case is comparable to the excited state lifetime of the hBN defects. We start from the inhibition factor, $\eta_{S}(I_{S},\tau_D,\tau_W)$, which is the ratio of fluorescence intensity with and without the STED laser, in the interval defined by the gate delay, $\tau_D$ and gate width, $\tau_W$, and can be approximated as: 

\begin{align} \label{Integral}
\eta_{S}(I_{S},\tau_D,\tau_W)=\int_{\tau_D}^{\tau_D+\tau_W}e^{-(k_{fl}+k_S)t}dt \bigg/\int_{\tau_D}^{\tau_D+\tau_W}e^{-k_{fl}t}dt  \tag{S2}
\end{align}

where $k_S = P_{S}\sigma_{S}\lambda_{S} \big/ A_{S}hc = \sigma_{S} I_{S}$ is the STED induced decay rate. Evaluating eq. \ref{Integral} leads to:

\begin{align} \label{inhib_factor}
\eta_{S}(I_{S},\tau_D,\tau_W)=\frac{1}{1+I_{S}/I_{sat}} \ \left[ \frac{\textrm{exp}(-\frac{\tau_D+\tau_W}{\tau_{fl}}\frac{I_{S}}{I_{sat}})-\textrm{exp}(\frac{\tau_W}{\tau_{fl}}-\frac{\tau_D}{\tau_{fl}}\frac{I_{S}}{I_{sat}})}{1-\textrm{exp}(\frac{\tau_W}{\tau_{fl}})} \right]  \tag{S3}
\end{align}
 where $I_{sat}=k_{fl}/\sigma_s$ is the saturation intensity, defined as the STED intensity for which the fluorescence is reduced by 50 \%. The effective PSF of the STED microscope, $h_{S}(r)$ arises from the multiplication of the Gaussian excitation profile by the inhibition factor, resulting in:

\begin{align} \label{hr}
h_{S}(r)=exp \left[ \frac{-r^24ln(2)}{d_c^2} \right]\frac{1}{1+S_r} \ \left[ \frac{exp(-\frac{\tau_D+\tau_W}{\tau_{fl}}S_r)-exp(\frac{\tau_W}{\tau_{fl}}-\frac{\tau_D}{\tau_{fl}}S_r)}{1-exp(\frac{\tau_W}{\tau_{fl}})} \right]  \tag{S4}
\end{align}

where $d_c$ is the FWHM of the Gaussian excitation profile and $S_r=P_{S}D(r)A_{S}/P_{sat}$. Here $P_{sat}$ is the STED saturation power and is used below as a fitting parameter.



\begin{figure}
  \includegraphics[width=\textwidth]{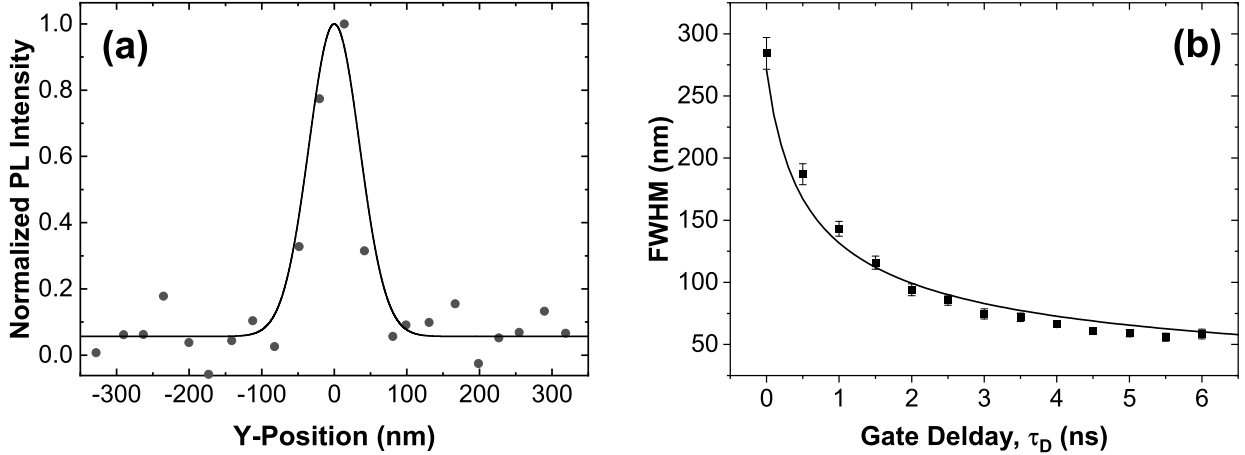}
  \caption{(a) Vertical intensity profile through the center of the STED image, shown in the Fig. 3(b) of the main text, with fit to the PSF calculated using eq. \ref{hr} and the parameters $\tau_D=3 \ \textrm{ns}$, $\tau_W = 0.5 \ \textrm{ns}$, $\tau_{fl}=1.7 \ \textrm{ns}$, $d_c = 450 \  \textrm{nm}$ and  $P_{sat} = 1.6 \ \textrm{mW}$. (b) Gaussian FWHM of the vertical intensity profile as a function of gate delay. The solid black line shows a fit to eq. \ref{hr} with the same parameters as (a).}
  \label{fgr_SI_fitting}
\end{figure}

Fig \ref{fgr_SI_fitting} (a) shows vertical intensity profiles through the center of the STED image shown in Fig. 3(a) of the main text. The solid line shows the point spread function calculated using eq. \ref{hr}, where the only fit parameter is $P_{sat} = 1.6  \textrm{mW}$ and which is in excellent agreement with the experimental data with FWHM of 61nm. The dependence of the FWHM of the STED PSF is plotted in Fig. \ref{fgr_SI_fitting}(b), along with the FWHM of the calculated PSF for $P_{sat} = 1.6  \textrm{mW}$ and an effective numerical aperture $\textrm{NA}_{eff} = 0.56$. From eq. \ref{hr} a prediction can be made for the resolution that could be obtained for higher NA. For an oil-immersion objective, where the NA can reach 1.45, the resolution can be as low as 6 nm.

\section{3. Additional STED cross section measurements}

We have measured the STED cross-section for a total of 10 different emitters. Fig. \ref{fgr_SI_example}(a)-(c) shows lifetime verses STED power for three of these defects. The mean STED cross section for the 10 emitters is $(5.5\pm3.2) \times 10^{-17} \mbox{ cm}^2$ with the highest $\sim 1 \times 10^{-16} \mbox{ cm}^2$ shown in \ref{fgr_SI_example}(a). From the same measurement, we also extract the average saturation power $P_{sat}=(14.7 \pm 9.5) \  \textrm{mW}$.

\begin{figure}
  \includegraphics[width=1\textwidth]{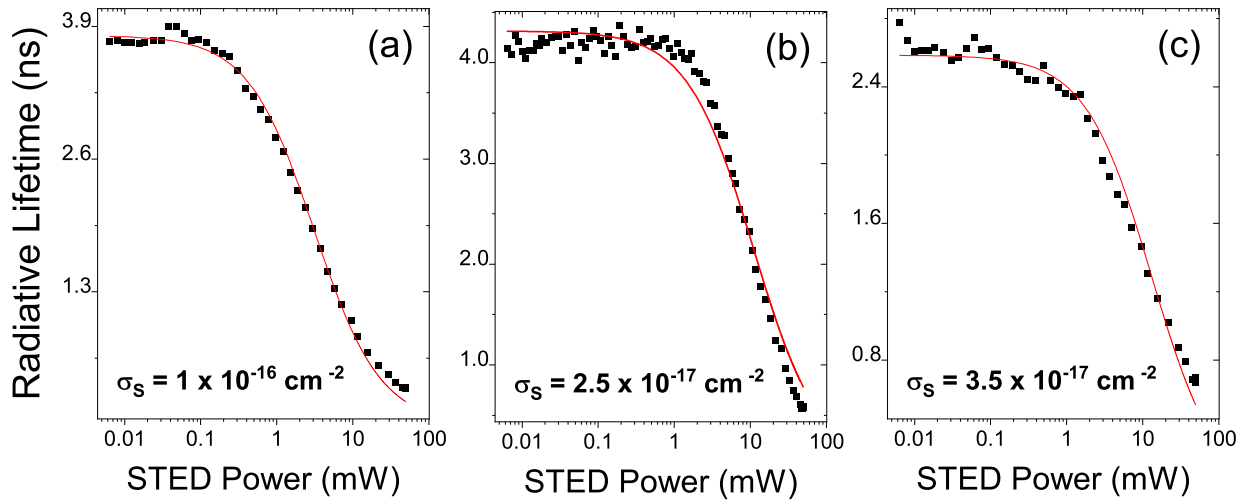}
  \caption{STED laser power dependence on radiative lifetime for 3 example color centers (a)-(c). $\sigma_{S}$ is calculated using eq. (1) from main text.}
  \label{fgr_SI_example}
\end{figure}

\end{suppinfo}

\bibliography{STED_ref}

\end{document}